\documentclass{elsart3}

\usepackage[dvips]{graphicx}
\usepackage{float}

\usepackage{amssymb}

\usepackage{url}

\newcommand{\celsius}{\ensuremath{^\circ\mathrm{C}}}

\begin{document}

\begin{frontmatter}


\title{Local Support Assembly of the ATLAS Pixel Detector}
\author{Karl-Heinz Becks},
\author{Peter Gerlach},
\author{Karl-Walter Glitza\corauthref{cor1}},
\ead{glitza@uni-wuppertal.de}
\author{Pascal Knebel},
\author{Peter M\"attig},
\author{Bernd Sanny},
\author{Sebastian Reuschel},
\author{Swetlana Springer},
\author{Bernd Witt}
\address{Bergische Universit\"at Wuppertal, Fachbereich C, 42097 Wuppertal, Germany}
\corauth[cor1]{Corresponding author. Tel.:+49-202-439-2738; fax: +49-202-439-2811}





\begin{abstract}
The barrel part of the ATLAS pixel detector will consist of 112 carbon-carbon structures called "staves" with 13 hybrid detector modules being glued on each stave. The demands on the glue joints are high, both in terms of mechanical precision and thermal contact. To achieve this precision a custom-made semi-automated mounting machine has been constructed in Wuppertal, which provides a precision in the order of tens of microns. As this is the last stage of the detector assembly providing an opportunity for stringent tests, a detailed procedure has been defined for assessing both mechanical and electrical properties. This note gives an overview of the procedure for affixation and tests, and summarizes the first results of the production.
\end{abstract}

\end{frontmatter}

\section{Introduction}
\label{sec:Introduction}

ATLAS will be a particle physics experiment at the future Large Hadron Collider (LHC), which is being built at CERN and is expected to start operation in 2007.

\begin{figure}[hbt]
	\centering
		\includegraphics[width=\columnwidth]{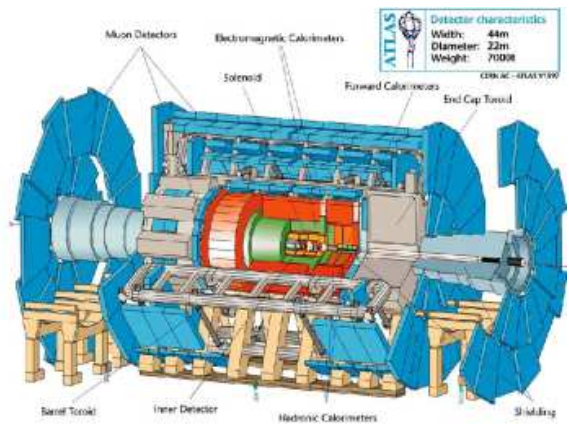}
	\caption{Overview of the ATLAS detector}
	\label{fig:ATLASdetector}
\end{figure}

The pixel detector is the innermost component of the ATLAS inner tracker. In the barrel the detector modules are mounted on staves, while the modules in the end caps are organized in disk sectors.

The pixel detector consists of 1774 detector modules (barrel: 1456 modules; discs: 318).

\begin{figure}[hbt]
	\centering
		\includegraphics[width=\columnwidth]{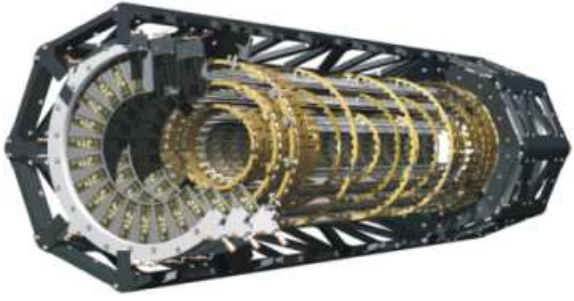}
	\caption{Overview of the PIXEL detector with three barrel layers and six disks}
	\label{fig:PIXELdetector}
\end{figure}

\nopagebreak
$\left.\right.$\\
\newpage

The most important components of a detector module are:
\begin{itemize}
	\item 46080 individual pixel sensors with size of $50 \mu m\ \times\ 400 \mu m$
	\item 16 front end read out chips
	\item 1 module controller chip
\end{itemize}

\section{Local Supports ("Staves")}
\label{sec:GlobalAndLocalSupports}

\begin{figure}[hbt]
	\centering
		\includegraphics[angle=0, scale=.45]{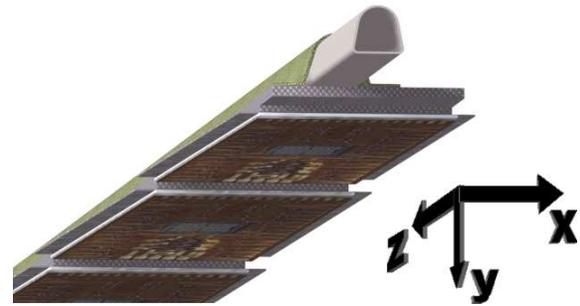}
	\caption{Details of the stave aluminium tube}
	\label{fig:ALtube}
\end{figure}

The pixel stave is the module local support unit of the barrel section of the pixel detector.
The components of a stave are the thermal management tile (TMT), the  stave carbon structure, and the 
stave cooling circuit. Additionally, every stave has two geometrical reference marks (ruby balls). The stave coordinate system is specified in Figure \ref{fig:ALtube}.

The TMT itself consists of two parts. Both parts have a shingled design with an angle of 1.1 degrees which are glued together.
 
As material for the TMTs, Carbon-Carbon (C-C) has been chosen. The reason for this is a thermal conductivity which is 10-100 times better than standard Carbon Fiber Reinforced Plastic (CFRP), even in the transverse direction to the fibres. It has excellent mechanical properties, stability, and transparency to particles.

The TMT is made of 1502 ZV 22 C-C material from SGL (Augsburg, Germany). The raw material is in the form of plates of about 6 mm thickness. The plates consist of 2-D roving fabric carbon fibres layers overlapped in a phenolic resin matrix, densified, and graphitized at high temperature to enhance the thermal properties. The raw TMTs are machined to the final stepping shape  with a CNC milling machine equipped with high speed spindle and diamond coated millers. 

The specific properties of the material are summarized in table \ref{tab:PropertiesOfSGLCc1502ZV22}.

\begin{figure}[hbt]
	\centering
		\includegraphics[width=\columnwidth]{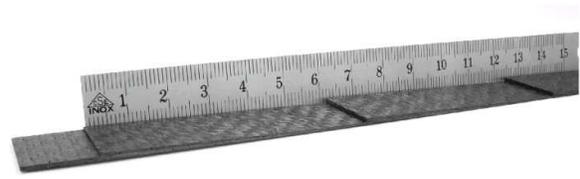}
	\caption{Shingled TMT with scale}
	\label{fig:TMT}
\end{figure}

\begin{table}[htb]
	\centering
		\begin{tabular}{|l|l|l|}
			\hline
			PROPERTY&UNIT&SPECIFICATION\\
			\hline
			Impurity&$ppm$&$< 100$\\
			\hline
			Density&$g/cm^3$&$1.76\pm0.04$\\
			\hline
			Porosity&\%&$13\pm2$\\
			\hline
			CTE - in plane&$10^{-6}K^{-1}$&$-0.5\pm0.2\ @ 300\celsius$\\
			\hline
			CTE - transverse&$10^{-6}K^{-1}$&$12\pm1\ @ 300\celsius$\\
			\hline
			Kt - in plane&$W/mK$&$250\pm10\ @ 20\celsius$\\
			\hline
			Kt - transverse&$W/mK$&$30\pm3\ @ 20\celsius$\\
			\hline
			Young's modulus - in plane&$GPa$&$120\pm20$\\
			\hline
		\end{tabular}
	\caption{Properties of  SGL cc 1502 ZV 22}
	\label{tab:PropertiesOfSGLCc1502ZV22}
\end{table}

The stave cooling circuit is made of a thin aluminum tube (see Figure \ref{fig:ALtube}), shaped to fit inside the inner cross section of the stave carbon structure.

The material chosen is 6060 aluminum alloy. This material shows good extrusion properties at small thickness.

The cooling system is based on the evaporation of fluorocarbon and provides an operating temperature of about $-6\celsius$.

The stave carbon structure ("omega profile") is made of three layers of unidirectional ultra high modulus (UHM) carbon fibre reinforced cyanate ester resin pre-preg (preimpregnated). The adopted lay-up (0-90-0), 0.3 mm thick, has been optimised through an extensive design and test program. The choice of the pre-preg material and of the lay-up with a central cross-layer has been made in order to match the longitudinal CTE to that of the impregnated C-C, thus minimizing distortions resulting during cool-down. The 13 modules are arranged one after the other along the stave axis and they are partially overlapped in order to achieve the required coverage in the axial direction. Thus, the surface of the stave in contact with the modules is stepped and the modules are arranged in a shingled layout.

However, C-C materials have two main technological drawbacks limiting their application range: porosity and difficulty to achieve complex and accurate geometries due to the high temperature manufacturing process.

To overcome the porosity of the C-C material, it was impregnated with resin such that infiltration by the thermal greases and carbon dust release could be avoided.

\section{Loading Machine}
\label{sec:LoadingMachine}
For the assembly of modules on staves a custom made module loading machine was developed and built in Wuppertal. The requirements of this machine are:
\begin{itemize}
	\item handling the modules with minimal stress
	\item positioning of modules on the stave with an accurancy better than 50 microns
	\item regular glue deposition\\
\end{itemize}                       
                       
To control the applied stress the bow of each module is measured before and after loading.

The main components of the module loading machine are the granite base of $1\ \times\ 2 m$ and a flatness of  $2 \mu m$ from Johan Fischer\footnotemark. On this base several linear guideways from Schneeberger\footnotemark type Monorail BM are mounted to allow movement of the central measurement unit, the microscope M 420 from Leica\footnotemark. The microscope itself is connected to OWIS\footnotemark micrometric tables and allows movements in all dimensions. The movements are controlled by Heidenhain\footnotemark sealed linear encoders. Heidenhain Digital Readouts type ND 760 are used for displaying and storing the position of the M 420 using a Personal Computer.
The module mounting head is connected to several linear guides, goniometer and rotary tables to reach any position. The module is fixed by vacuum to the mounting head and its position is always monitored by the microscope.

For the deposition of the glue a computer controlled dispenser from EFD\footnotemark type 1502 is used. The assembly time of each module is about 1 hour, the curing time of the glue is 2 hours. This leads to a production rate of 1 stave per week.
\addtocounter{footnote}{-5}
\footnotetext{Johannes Fischer, Aschaffenburg}
\stepcounter{footnote}
\footnotetext{Schneeberger, H\"ofen, Enz}
\stepcounter{footnote}
\footnotetext{Leica Microsystems, Bensheim}
\stepcounter{footnote}
\footnotetext{OWIS, Staufen}
\stepcounter{footnote}
\footnotetext{Heidenhain, Traunreut}
\stepcounter{footnote}
\footnotetext{EFD/GLT, Pforzheim}

\begin{figure}[hbt]
	\centering
		\includegraphics[width=\columnwidth]{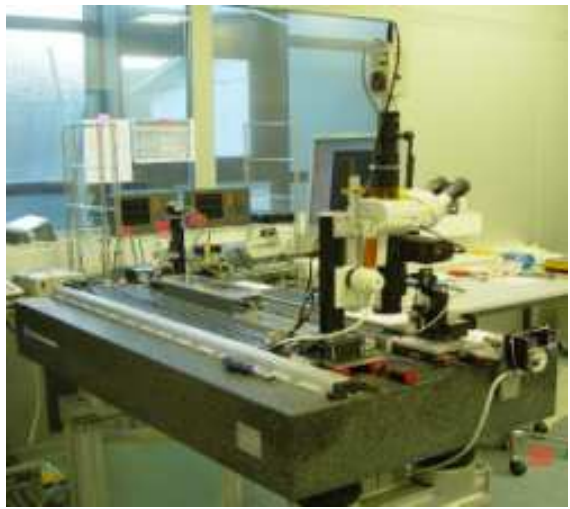}
	\caption{Wuppertal loading machine with granite table and microscope positioning system}
	\label{fig:loadingmachine}
\end{figure}

$\left.\right.$\\[0.5cm]

\section{Results}
\label{sec:Results}

\begin{figure*}[!ht]
	\centering
		\includegraphics[width=1.00\textwidth]{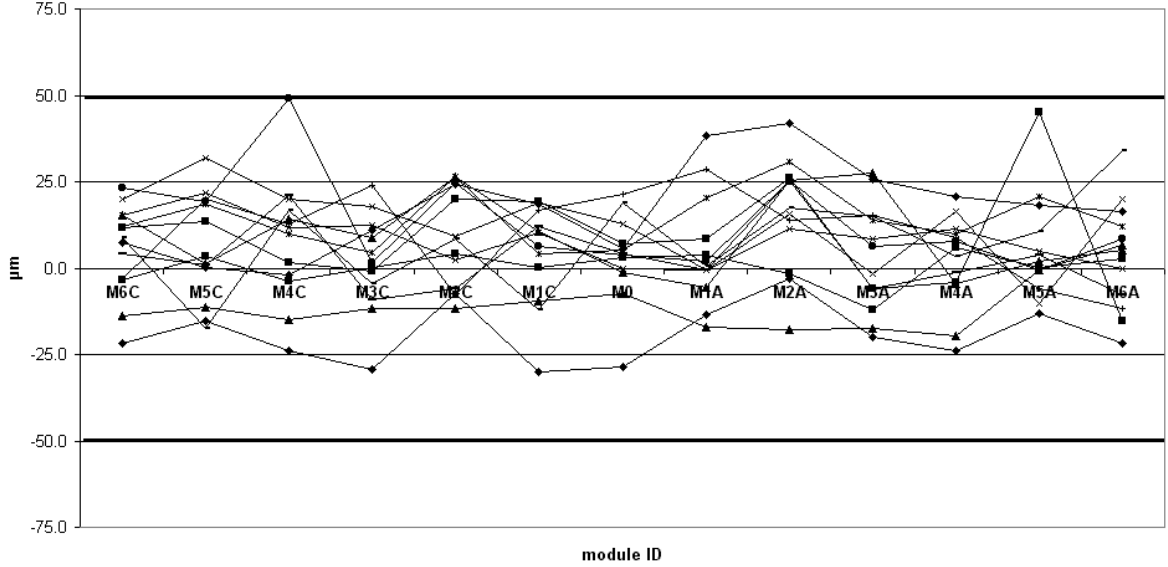}
	\caption{Z-deviation for each module for several staves}
	\label{fig:z-deviation_modwise}
\end{figure*}

The x, y, and z position of the glued modules of each stave is controlled with respect to the stave ruby balls. These provide the reference system for the module position and allow one to assess the accuracy of the loading procedure.

As a typical result the z position measurement for 13 staves is provided in Figure \ref{fig:z-deviation_modwise}. For each stave the deviation from the nominal position is shown for each of the 13 modules. The stated module ID's are equivalent to defined locations on the stave. The tolerances are $\pm50\mu m$ and are indicated in the figure by thicker lines.

\begin{figure}[hbt]
	\centering
		\includegraphics[angle=0, scale=0.35]{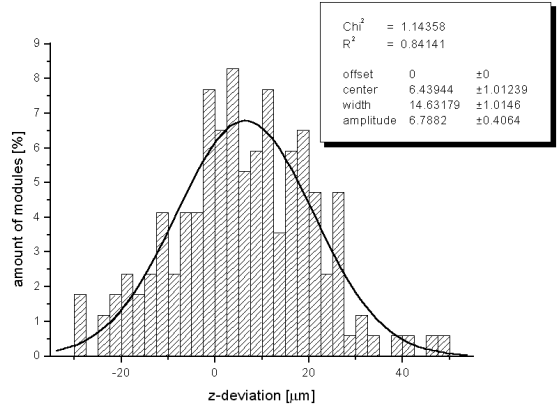}
	\caption{Distribution of z-deviation}
	\label{fig:z-deviation_distribution}
\end{figure}

The plot shows that the accuracy of the z positioning is always within the tolerances of $\pm50 \mu m$. In Figure \ref{fig:z-deviation_distribution}, the distribution of the z-deviation is given and demonstrates that 50\% of the modules are glued with an accuracy which is even better than $15 \mu m$. One can also see that there is a systematic shift of 6-7$\mu m$.

\begin{figure*}[!ht]
	\centering
		\includegraphics[width=1.00\textwidth]{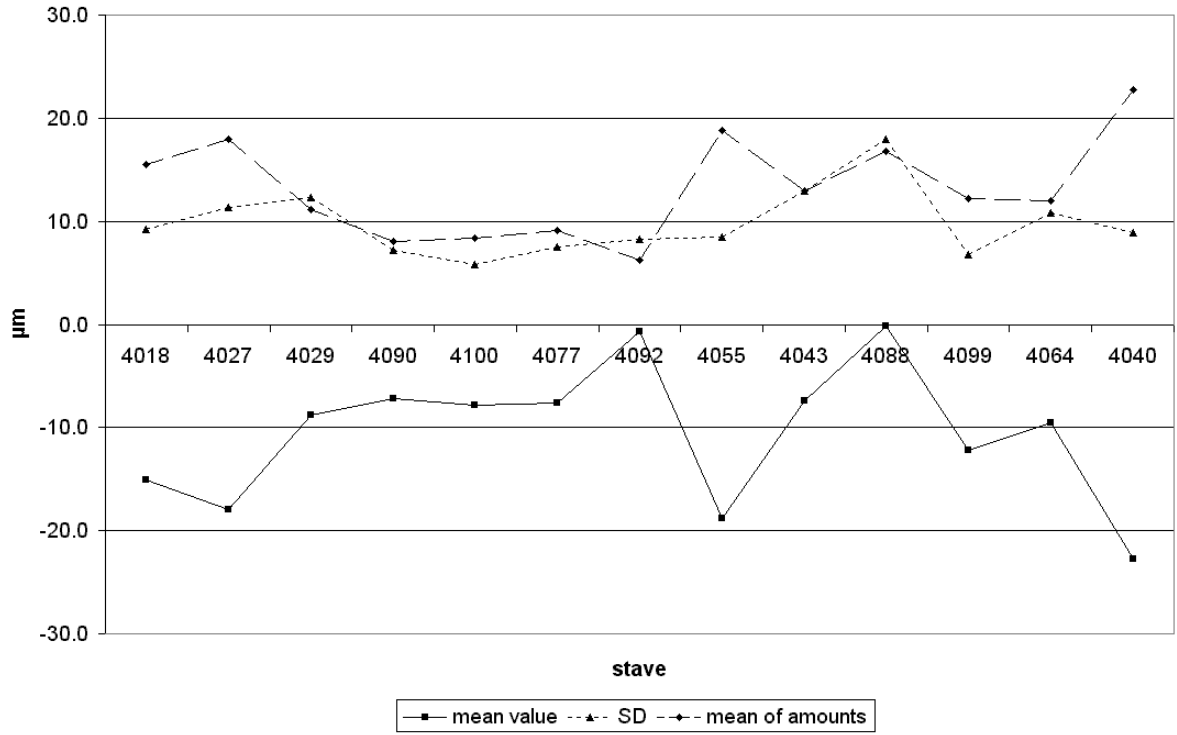}
	\caption{Difference between bow before and after loading}
	\label{fig:bow}
\end{figure*}

As mentioned previously, the difference between the bow of the module before and after loading is an indicator as to whether stress has been applied during the loading procedure. Figure \ref{fig:bow} shows the mean bow difference (mean value), averaged over all 13 modules of one stave, as well as the Standard Deviation (SD) and the average of the absolute values (mean of amounts).

One can see, that the bow difference is always better than $25\mu m$. This is interpreted to show that a minimum of stress is being applied to the modules during the loading procedure.\\[1.166cm]

\section{Conclusion}
\label{sec:Conclusion}

A module loading machine for the ATLAS pixel detector has been successfully realized in Wuppertal. All requirements of the mounting precision are fulfilled. The position of the modules after loading are well within the tolerances. The applied stress during loading is negligible.

\setcounter{secnumdepth}{0}
\section{Acknowledgement}
\label{sec:Acknowledgement}

This work has been supported by the \emph{Bundes-ministerium f\"ur Bildung, Wissenschaft, Forschung und Technologie} (BMBF) under grant number\linebreak 05 HA4PX1/9.

\newpage



\end{document}